# Early-Time Reshaping of Laser-Induced Plasma Profiles

*Ivan Ostrovsky[1,2], Gilad Hurvitz[3], Soumitra Hazra[1,2], and Sharly Fleischer[1,2]*

[1] School of Chemistry, Raymond and Beverly Sackler Faculty of Exact Sciences, Tel Aviv University, Tel Aviv 6997801, Israel.
[2] Tel Aviv University Center for Light-Matter Interaction, Tel Aviv University, Tel Aviv 6997801, Israel.
[3] Applied Physics Division, Soreq NRC, Yavne 8180000, Israel

## Abstract

Laser-induced plasmas are primarily characterized by their temporal density decay. Using time-resolved transverse optical diffractometry, we reveal their concurrent spatial evolution, demonstrating pronounced broadening and flattening during the first 200 ps following ionization. This non-self-similar evolution arises from the local density dependence of electron-ion recombination, without particle transport. Exploiting the continuum of initial densities within a single plasma disc, we experimentally reconstruct an effective, time-dependent local recombination law over a broad plasma density range. Applied locally to independently measured initial plasma profiles, this kinetic map quantitatively predicts their subsequent evolution. These results are relevant to transient diffractive optics, plasma-based optical elements and waveguides, and the development of laser-induced plasmas as platforms for gas-phase THz plasmonics.

## Introduction

Laser-induced air plasmas constitute one of the most versatile manifestations of ultrafast nonlinear light–matter interactions spanning a broad range of optical and atmospheric applications, including laser-induced breakdown spectroscopy [1–4], filament-guided electrical discharges [5–8], remote sensing [9,10], air lasing [11–15] and optical wave-guiding [16] among many others [6–8,17–32]. In all these applications, the plasma created immediately following laser ionization establishes the initial conditions for its subsequent evolution. Understanding the physical processes governing this early stage is therefore essential for predicting and ultimately controlling laser-induced plasma dynamics for its practical utilization.

The spatial and temporal dynamics of laser-induced air plasmas have been the subject of extensive experimental and theoretical investigations [8,33–35]. Previous studies have characterized the evolution of the electron density, electron temperature, plasma lifetime, and plasma chemistry using interferometric [27,32–39], diffractometric [30,40–44], spectroscopic [14,15,45–53], electrical [6,54], and optical diagnostics [27,55–59]. Despite these advances, the evolution of the plasma profile during the first few hundred picoseconds following laser ionization has received comparatively little attention. Most studies have focused on the temporal decay of the plasma density, implicitly treating the plasma profile as a secondary quantity.

If the plasma profile retained its shape during the decay, its evolution would be fully described by the temporal decrease of the plasma density. Any deviation from self-similar evolution, however, indicates that different regions of the plasma decay at different rates and therefore provides direct insight into the underlying plasma kinetics. Quantifying such reshaping has received little attention despite its potential to reveal the local dynamics governing the earliest stages of plasma evolution.

In this work, we combine time-resolved transverse optical diffractometry with SuperGaussian (SG) distribution analysis to investigate the evolution of femtosecond laser-induced air plasmas during the first 200 ps following optical breakdown. From measurements of a single plasma disc, we experimentally determine the local time-dependent recombination law over a broad range of initial plasma densities and subsequently use it to predict the evolution of several independent plasma profiles.

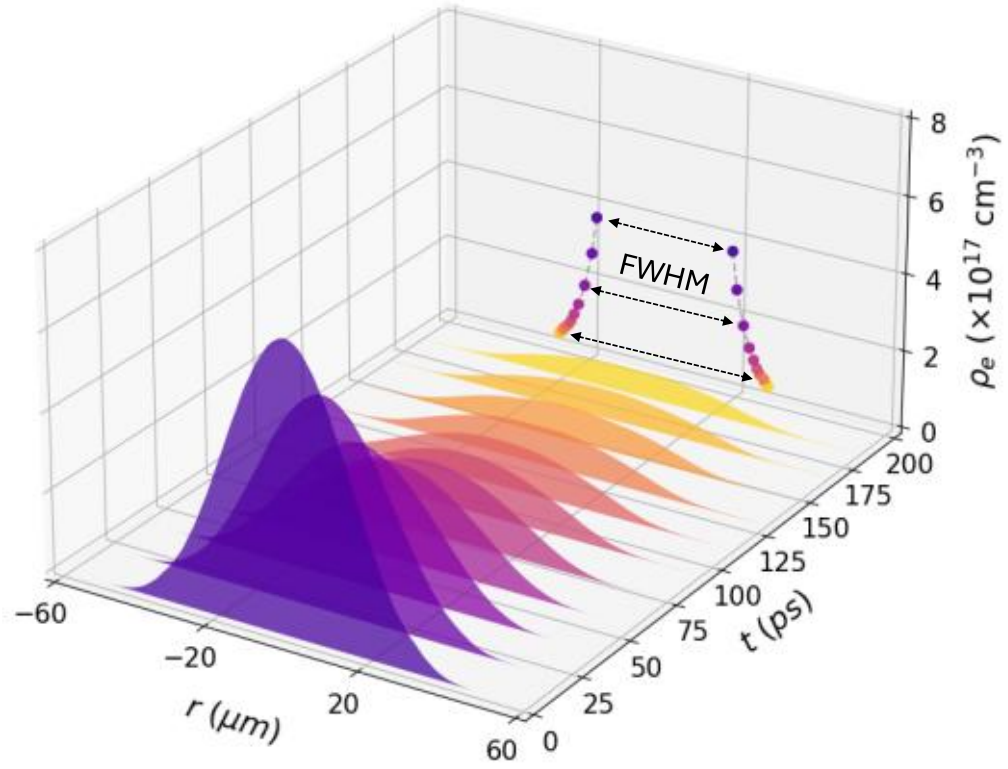

**Figure 1.** Time evolution of a transverse plasma density profile ("plasma disc") during the first 200 ps following femtosecond laser-induced breakdown in air. While the peak electron density decreases with time, the full width at half maximum (FWHM), projected onto the rear plane, increases from ≈38 μm to ≈58 μm, accompanied by a gradual flattening of the plasma profile.

Figure 1 shows the temporal evolution of the transverse plasma density profile at a fixed longitudinal position within the plasma column during the first 200 ps following laser-induced breakdown. Throughout this manuscript, we refer to such a transverse slice as a *plasma disc*. The radially resolved plasma density is reconstructed at several pump-probe delays from time-resolved transverse optical diffractograms [60]. As expected, the peak plasma density decreases monotonically with time. Simultaneously, however, the plasma profile exhibits a continuous increase in its full width at half maximum (FWHM), from approximately 38 μm to 58 μm (projected onto the rear plane of Fig. 1), indicating that the plasma does not undergo self-similar decay. In addition to this broadening (see SI Sec. 2), the profiles exhibit gradual flattening, the quantitative characterization of which is given below.

The increase in plasma width might naturally be attributed to radial particle transport through ambipolar diffusion. For laser-induced air plasmas at atmospheric pressure, the ambipolar diffusion coefficient is approximately $D_a \sim 10^{-4} m^2/s$, yielding a characteristic transport length $L_D = \sqrt{2D_a t} < 0.2\ \mu m$ during the first 200 ps [57]. This distance is more than two orders of magnitude smaller than the measured plasma width ($\sigma \approx 15 - 20\ \mu m$) and far below the observed increase in FWHM. Ambipolar transport can therefore be neglected during the time-interval investigated here. In the absence of significant particle transport, the plasma evolution is governed by the local electro-ion recombination, described by the two-body recombination equation, $\frac{\partial \rho}{\partial t} = -\beta \rho^2$ (eq.1)
where ρ is the electron density and β is the electron-ion recombination coefficient [32,43,56,57,61].

The quadratic dependence of Eq. (1) on the local plasma density implies that the plasma cannot, in general, decay self-similarly. Instead, the dense plasma core recombines more rapidly than the lower-density wings, producing a spatially inhomogeneous local decay that gradually broadens and flattens the plasma density profile, even in the absence of particle transport. Throughout this work, we use $\beta$ to describe an effective local two-body electron-loss coefficient that incorporates the evolving recombination kinetics over the investigated density and temporal range.

To quantify the plasma profile evolution, we parameterize the measured radial density distribution by a supergaussian (SG) model, $\rho_e(r,t) = \rho_0 \cdot \exp\left[-\left(\frac{r}{\sqrt{2}\sigma}\right)^{2\gamma}\right]$, where $\rho_0$ is the central electron density, $\sigma$ is the width parameter, and $\gamma$ is the SG order. While $\rho_0$ characterizes the plasma decay, the parameters $\sigma$ and $\gamma$ quantify the accompanying broadening and flattening of the plasma profile respectively. The temporal evolution of these parameters quantitatively characterizes the reshaping of the plasma. Experimentally, $\rho_e(r,t)$, is reconstructed from transverse optical diffractometry using a thin ($\approx 20\mu m$ longitudinal resolution) near-IR pulse and fitted to the SG model [60].

## Experimental system

Plasma columns were generated by a $\approx 120$ fs near-IR pulse with varying pulse energies between $0.1$ and $1$ mJ, focused by a $f = 500$ mm lens in ambient laboratory atmosphere. Transverse plasma density distributions were obtained using the optical diffractometry technique introduced in Ref. [60]. An ultrafast near-IR probe pulse ($\approx 120$ fs) traversed the plasma column perpendicular to its propagation direction, providing a longitudinal resolution of $\approx 20\mu m$. Diffraction patterns were acquired over the first $200$ ps following plasma formation for a range of pulse energies and longitudinal positions along the plasma column. The optical phase imprinted on the probe pulse was retrieved using the analytical phase-reconstruction method of Ref. [60]. The resulting transverse phase profile was fitted to a SG plasma distribution, yielding the central electron density ($\rho_0$), width parameter ($\sigma$), and a SG order ($\gamma$). Measurements were performed over a longitudinal range of $\pm 5$ mm around the focal plane of the pump. Representative diffraction patterns, retrieved phase profiles, and SG fits are provided in Supplementary Information (SI) Section 1.

## Recombination kinetics at varying initial plasma densities

To examine the recombination kinetics over a broad range of plasma conditions, we measured the temporal decay of the electron density at the center (r=0) of plasma discs exhibiting different initial electron densities. Representative decay curves are shown in Fig. 2a, where the symbols denote the measured central density, $\rho_0(t)$, normalized to its initial value, $\rho_0(0)$. The dashed-dotted curves are fits to the two-body recombination model of Eq. (1), assuming a constant recombination coefficient ($\beta$),

$$\rho_0(t) = \left(\frac{1}{\rho_0(0)} + \beta t\right)^{-1} \quad \text{(eq.2)}$$

The extracted recombination coefficients, indicated in Fig. 2a, are found to decrease systematically with increasing initial plasma density. Figure 2b summarizes the fitted $\beta$ values obtained from plasma discs spanning a broad range of longitudinal positions, pulse energies, and initial profile parameters, revealing a systematic decrease of $\beta$ with increasing $\rho_0(0)$. This trend is attributed to the higher plasma temperatures associated with regions exposed to greater laser intensity: while higher local intensities generate larger initial electron densities, they also produce hotter plasmas, for which electron-ion recombination is known to be less efficient [27,30,42,62–65].
Despite the reduction in $\beta$, the plasma lifetime decreases with increasing initial density, where $t_{1/2} = (\beta\rho_0)^{-1}$ denotes the time required for the plasma density to decay to half its initial value (Eq.2). As shown in Fig. 2a, this time is 30 ps for the densest plasma (purple curve), compared with 138 ps for the least dense plasma (yellow curve). Hence, over the investigated density range, the increase in $\rho_0$ dominates over the reduction in $\beta$, leading to progressively shorter plasma lifetimes.

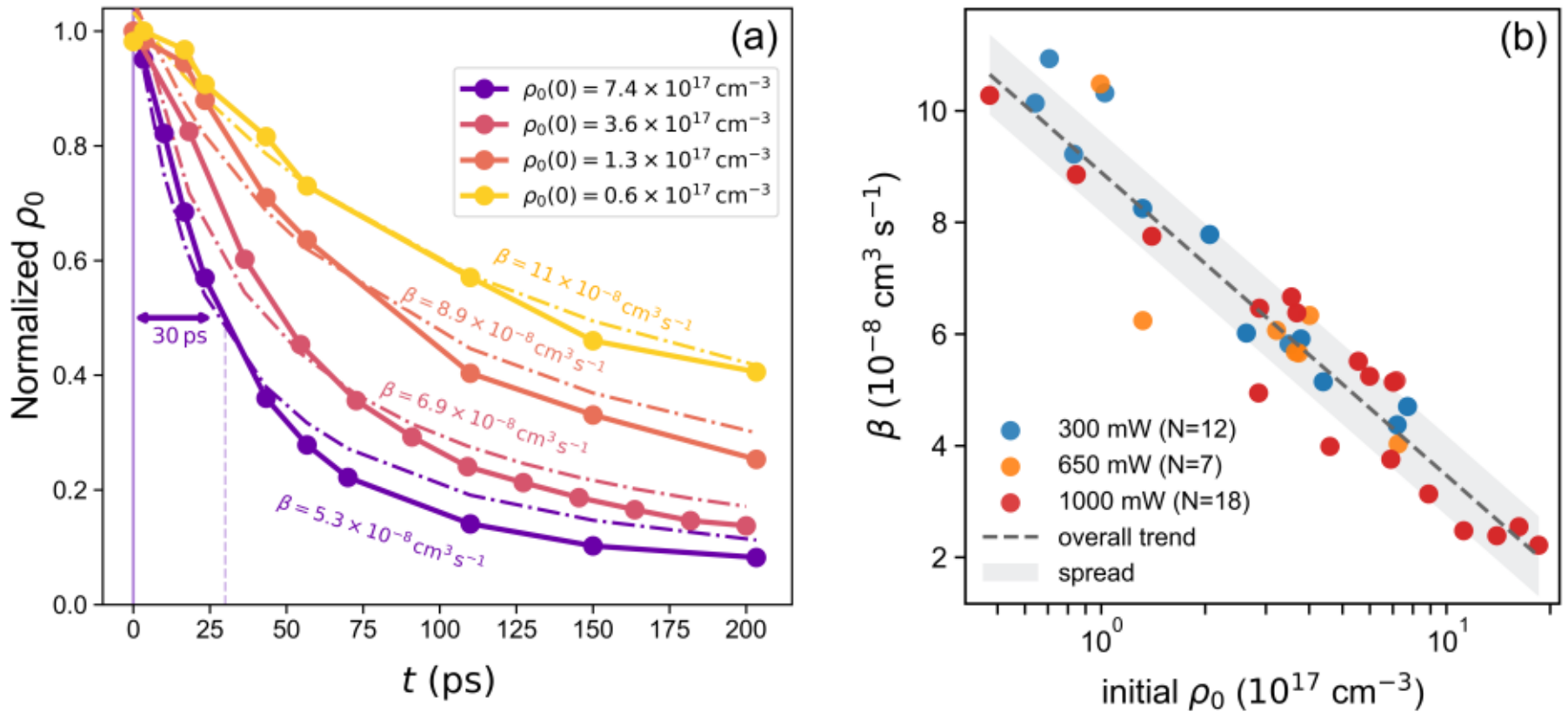

**Figure 2**. (a) Normalized plasma-density decay curves plotted versus time. Each trace was fitted with the standard two-body recombination model [Eq. (2)], treating the recombination coefficient $\beta$ as constant in time. Across all cases, the fits systematically fail to capture the late-time behavior, indicating $\beta$ is a function of time. (b) Fitted $\beta$ values versus the initial peak electron density, grouped by pump energy. The dashed line shows the overall log-linear trend, and the grey band represents $\pm 1$ residual standard deviation about the trend. The systematic decrease in $\beta$ indicates a reduction of the effective recombination coefficient with increasing initial density.

A closer inspection of Fig.2a reveals a systematic deviation between the experimental data and the fit with constant $\beta$ (eq.2). Specifically, the fitted curves consistently overestimate the measured plasma density at delays exceeding $\sim 100$ps, indicating that the measured plasma decays progressively faster with time. A natural explanation is provided by the temporal evolution of the electron temperature. Immediately following ionization, the plasma is hot, and the electron temperature subsequently decreases through collisional cooling [32]. Since the recombination coefficient is known to increase as the electron temperature decreases [27,30,42,62–65], $\beta$ cannot remain constant throughout the plasma evolution.
The experimentally observed acceleration of the decay is therefore consistent with a gradual increase of $\beta$ with time. Introducing an additional background-density parameter, as proposed in Refs. [21,23,60,61], did not improve the quality of the fit (see SI Sec. 3).
These observations motivate treating $\beta$ as a time-dependent quantity, $\beta(t)$. We parametrize $\beta(t)$ by the plasma's initial conditions rather than by the instantaneous electron density. This choice is supported by comparisons of plasma discs with similar initial peak densities but substantially different radial distributions, characterized by different $\sigma$ and $\gamma$. Despite their distinct geometries, these plasmas exhibit similar decay curves (see SI Sec. 3), indicating that the recombination dynamics are governed primarily by the initial plasma density. Since the initial plasma density is correlated with the laser intensity and the corresponding electron temperature, we express the recombination law as $\beta(t; \rho(0))$, which provides the basis of the predictive model developed below.

## Experimental determination of the local recombination law

To determine the temporal evolution of $\beta$, we first consider a plasma disc characterized by a specific initial electron density, $\rho_0(0)$. We acquired densely sampled diffraction measurements from a single plasma disc throughout the first 200 ps following plasma creation and extract the electron density at the center of the disc vs. time, $\rho_0(\mathrm{t})$, shown in Fig.3a. To determine the temporal evolution of $\beta$ we perform a sliding-window analysis by fitting equation (2) to consecutive nine data point segments ($\sim 20-30$ ps duration). Within each segment (exemplified by the red points in Fig.3a) we treat $\beta$ as constant. The fitted $\beta$ value is assigned to the temporal center of the corresponding window, yielding an estimate $\beta(t_{segment})$. Repeating the procedure over the entire temporal trace produces the evolution of the effective recombination coefficient $\beta(t)$, shown in Fig. 3(b).

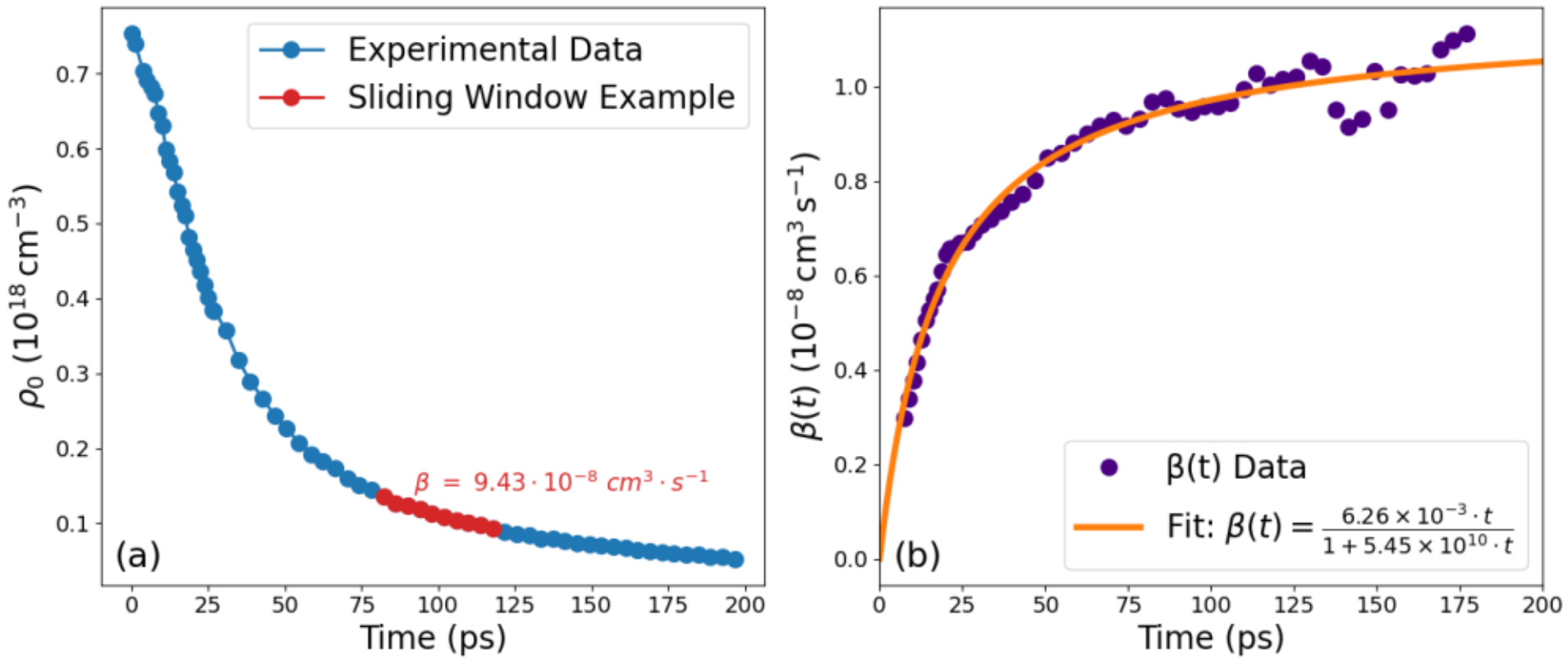

**Figure 3.** Sliding-window extraction of the time-dependent effective recombination coefficient. (a) Densely sampled temporal decay of the peak electron density, $\rho_0(t)$. The red points highlight a representative short-time window used for a local two-body recombination fit, with the extracted $\beta$ value indicated. (b) $\beta(t)$ obtained by translating the window across the full decay. The solid orange curve is a fit to the rational model $\beta(t) = at/(1 + bt)$.

The extracted β exhibits a rapid increase during the first 50 ps, followed by a markedly slower evolution toward a quasi-stationary value, consistent with the fast electron cooling in the vicinity of plasma formation. To provide a compact parametrization of the measured dynamics, the extracted coefficients, $\beta(\mathrm{t_{segment}})$, are fitted by the empirical rational function, $\beta(\mathrm{t}) = \frac{\mathrm{at}}{1+\mathrm{bt}}$ , where a and b are free parameters. Within our experimental window, this function accurately captures the observed two-stage evolution of the plasma: an initial rapid increase during the first few tens of ps, followed by a considerably slower evolution at later times.

We emphasize that the rational form is adopted solely as an empirical parametrization over the 10-200 ps interval investigated here. It is not intended to describe the asymptotic plasma behavior, for which the imposed limits $\beta(0) = 0$ and $\beta(\infty) = \mathrm{a/b}$ have no particular physical significance. Nevertheless, across the wide range of plasma conditions investigated, it provides an accurate and robust description of the measured recombination dynamics.

Importantly, the temporal evolution of β depends parametrically on the initial plasma density. Regions of higher initial density are generated by higher local laser intensities and therefore originate from hotter plasmas, which are characterized by lower initial recombination coefficients. Likewise, the subsequent increase of β reflects plasma cooling dynamics and therefore also depends on the initial plasma conditions. In the following section, we exploit this dependence to construct the local recombination law, $\beta(\mathrm{t}; \rho(0))$, over the full range of initial plasma densities.

## Constructing the recombination map from a single plasma disc

The time-resolved evolution of a single plasma disc contains the recombination trajectories of all resolved initial densities represented across its SG radial profile. Owing to the negligible transport during the first 200 ps, each radial position can be treated as a locally evolving plasma element characterized by its initial density, $\rho_{in}(r) = \rho(r, t = 0)$, covering the range $0 \leq \rho_{in} \leq \rho_{max}$.

The rational temporal dependence established in the previous section is applied separately to each initial local density: $\beta(t) = \frac{a(\rho_{in})t}{1+b(\rho_{in})t}$. Substitution into the recombination equation gives the explicit density evolution:

$$(3)\ \rho(t; \rho_{in}) = \left[\frac{1}{\rho_{in}} + \frac{a_{in}}{b_{in}^2}\left(b_{in}t - ln(1 + b_{in}t)\right)\right]^{-1} \qquad (3)$$

where $a_{in} = a(\rho_{in})$ and $b_{in} = b(\rho_{in})$. Fitting Eq. (3) to the trajectory extracted at each resolved radial position determines $a(\rho_{in})$ and $b(\rho_{in})$, thereby reconstructing the empirical local recombination map $\beta(t; \rho_{in})$.

Note that the dense temporal sampling required for the sliding-window analysis of Fig.3 established the temporal functional form of $\beta(t)$. Equipped with the latter we can construct the local recombination map with a substantially sparser series of SG profiles since each local trajectory is described by the rational function fitted by only two free parameters. Using this strategy, illustrated schematically in Fig.4, a single plasma disc, sampled at several delays provides a continuum of simultaneous recombination trajectories from which the complete density-dependent recombination law is obtained. Naturally, the latter spans the experimentally sampled temporal and initial density region $[t, \rho_{in}]$. For detailed analysis of the construction of $\beta(t; \rho_{in})$ we refer the readers to SI sec. 5.

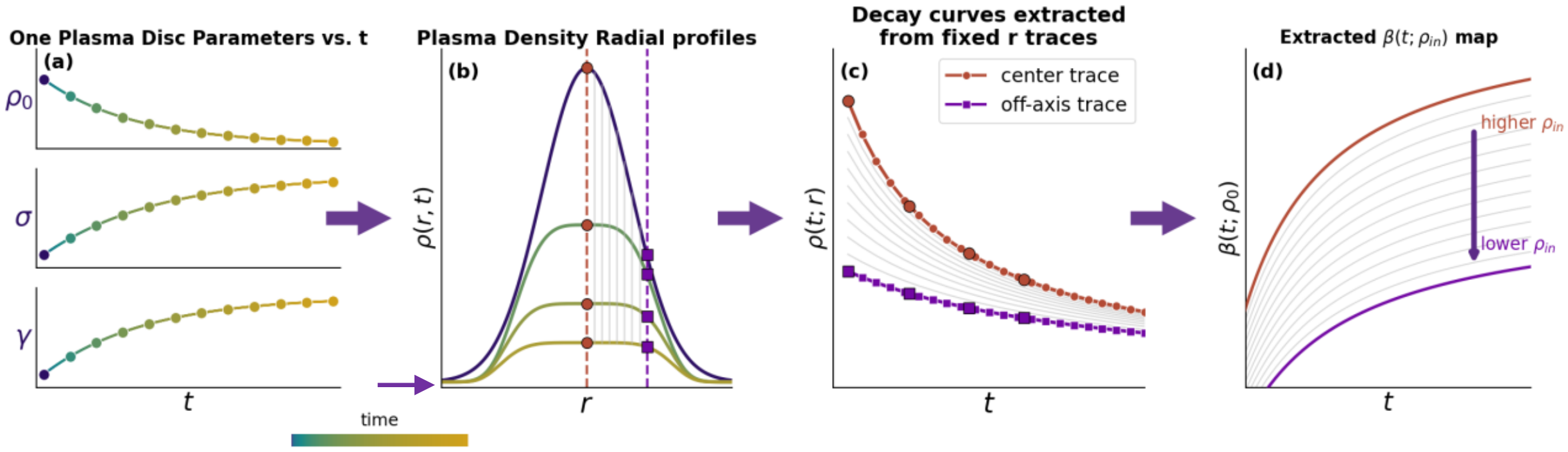

**Figure 4.** Reconstruction of the density-dependent recombination law from a single time-resolved plasma disc. (a) Temporal evolution of the fitted supergaussian parameters $\rho_0$, $\sigma$, and $\gamma$ at a fixed longitudinal position. Colors progress from dark at early times to yellow at later times. (b) Corresponding radial electron-density profiles. Each radial position represents a plasma element with a different initial local density. (c) Local decay trajectories extracted at fixed radii and fitted using Eq. (3); the center and a representative off-axis trajectory are highlighted. (d) The fitted trajectories collectively construct the empirical recombination map $\beta(t; \rho_{in})$ over the range of initial densities contained within the plasma disc.

## Predicting plasma reshaping from the local recombination law

Equipped with the local recombination law $\beta(t; \rho_{in})$ obtained from the analysis of a from a single plasma disc (Fig.4d), we now examine the model's capability to predict the evolution of independent plasma profiles, spanning a broad range of initial peak densities and SG parameters. For each independent plasma disc (shown in Fig.5), we experimentally measured the initial profile at t=0, serving as the sole input to the model, and predicted its evolution up to 200ps.

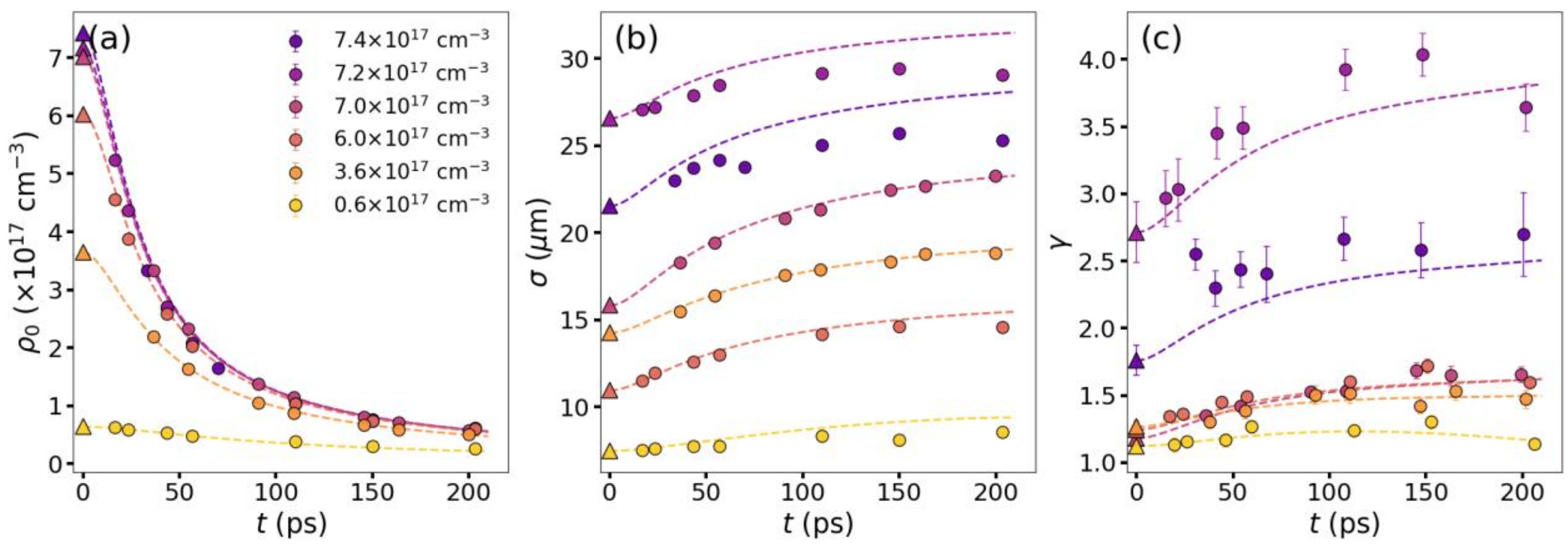

**Figure 5.** Prediction of plasma-profile evolution using the local recombination map $\beta(t;\rho_{in})$. Temporal evolution of (a) the peak electron density $\rho_0$, (b) the SG width $\sigma$, and (c) the SG order $\gamma$ for 6 plasma discs spanning a broad range of initial densities and profile parameters. Color-coded triangles mark the experimentally measured initial SG parameters (and t=0) that seed the recombination law model. The predicted temporal evolution of the SG parameters is depicted by the color-coded dashed curves. The circles represent independent plasma measurements, in agreement with the SG values predicted by the recombination law, hence validate the predictive power of the model.

Figure 5 compares the predicted evolution (dashed curves) with independent measurements (circles).Six plasma discs, located at arbitrarily chosen longitudinal positions and induced by varying pulse energies are examined, spanning a broad range of plasma parameters. Their initial SG parameters ($\rho_0(0)$, $\sigma(0)$ and $\gamma(0)$), depicted by triangular signs, are obtained experimentally and define the initial plasma distribution. Each radial element drawn from the initial distribution is propagated in time according to its corresponding recombination trajectory, $\beta(t;\rho_{in})$, to obtain $\rho(r,t)$. The latter is fitted to a SG distribution at selected timed to yield the predicted evolution of the SG parameters shown by the dashed curves. To validate the predicted evolution we measured the plasma discs at several selected times and extracted their SG parameters shown by the circular data points.

As readily observed, the model reproduces the evolution of all three observables over the first 200 ps with satisfactory accuracy. In particular, it captures the increase in plasma width (Fig.5b) and the evolution of the SG order (Fig. 5c) across the full range of initial plasma conditions. The reshaping dynamics is found to depend strongly on the initial on-axis density: plasma discs with higher initial densities $\rho_0(0)$ broaden and flatten more rapidly than low initial-density discs. For example, in the low-density case, $\rho_0(0) = 6 \times 10^{16} cm^{-3}$ (yellow), $\gamma$ increases only modestly, from $\approx 1.1$ to $\approx 1.3$ over 200 ps whereas in the high-density case, $\rho_0(0) = 7.2 \times 10^{17} cm^{-3}$ (blue), $\gamma$ rises from $\approx 1.8$ to $\approx 2.5$ over the same interval. A similar trend is observed in $\sigma(t)$, consistent with the higher initial electron temperatures of denser plasma discs. Consistent with these findings, a time-dependent flattening of the plasma channel has been mentioned in [27,32].

While the uncertainties in $\rho_0$ and $\sigma$ are smaller than the plot markers and are therefore omitted in Figs. 5(a) and 5(b), the larger uncertainties in $\gamma$ reflect its greater sensitivity to variations in the profile shape.

This stems from the inherent sensitivity of the SG order parameter, whereas, as $\gamma$ increases, the SG distribution becomes increasingly self-similar. For example, while deciphering $\gamma = 1.3$ from $\gamma = 1.6$ is readily accomplished, deciphering $\gamma = 2$ from $\gamma = 3$ is more challenging. This is manifested in the deteriorating agreement between the predicted and experimental data at the highest initial densities, correlated with higher $\gamma$ values. A detailed error analysis is provided in SI Sec. 4.

Importantly, while the results of Fig.5 verify the ability of the model to predict the early time reshaping of the plasma discs, the parameter range at is confined within the plasma densities over which the recombination map was constructed and should not be extrapolated beyond this range. This is because at initial peak densities exceeding $\sim 5 \times 10^{18}\, cm^{-3}$, our measurements reveal a qualitatively different trend in the recombination dynamics, whereby denser plasmas decay more slowly rather than more rapidly, as observed before. These high-density dynamics remain beyond the scope of this paper.

## Conclusions

We have investigated the early-time evolution of laser-induced plasma discs during the first 200 ps following their generation. The observed broadening and flattening of the plasma, parametrized by a super-Gaussian density distribution, emerge naturally from local density-dependent recombination, without invoking particle transport or diffusion. The extent and dynamics of the plasma reshaping depend primarily on the initial local density and can be predicted by applying the extracted recombination law locally across the plasma. This recombination-driven spatial evolution has important implications for various practical applications, including plasma-based optical elements, plasma holography and transient diffractive optics, plasma-based waveguides. We further note that the plasma frequencies associated with the densities investigated here lie within the THz frequency range. More broadly, understanding the early-time evolution of the plasma density distribution may provide an important foundation for exploiting laser-induced plasmas as dynamic platforms for gas-phase THz plasmonics.

- The authors acknowledge the support of Pazy Foundation grant no. 5100057425 and the Israeli Science Foundation grant no. 1856/22.